# EUSO@TurLab project in view of Mini-EUSO and EUSO-SPB2 missions

**H. Miyamoto[1,2*], M. Bertaina[1,2], D. Barghini[1,2,12], M. Battisti[1,2], A. Belov[3], F. Bisconti[1,2], S. Blin-Bondil[4], K. Bolmgren[8], G. Cambie[6,7], F. Capel[8], R. Caruso[9,10] M. Casolino[6,7,11], I. Churilo[13], G. Contino[9,10], G. Cotto[1,2], T. Ebisuzaki[11], F. Fenu[1,2], C. Fuglesang[8], A. Golzio[1,2], P. Gorodetzky[4], F. Kajino[18], P. Klimov[3], M. Manfrin[2] L. Marcelli[6,7], M. Marengo[1] W. Marszał[14], M. Mignone[1], E. Parizot[4], P. Picozza[6,7], L.W. Piotrowski[11], Z. Plebaniak[14], G. Prévôt[4], E. Reali[7], M. Ricci[17], N. Sakaki[11], K. Shinozaki[14], G. Suino[1,2], J. Szabelski[14], Y. Takizawa[11], A. Youssef[2] on behalf of the JEM-EUSO Collaboration**

(a complete list of authors can be found at the end of the proceedings)

*[1]INFN Turin, Italy, [2]University of Turin, Department of Physics, Italy, [3]SINP, Lomonosov Moscow State University, Moscow, Russia., [4]APC, Univ Paris Diderot, CNRS/IN2P3, France, [5]INFN Bari, Italy, [6]INFN Tor Vergata, Italy, [7]University of Roma Tor Vergata, Italy, [8]KTH Royal Institute of Techinology, Stockholm Sweden, [9]University of Catania, Italy, [10]INFN Catania, Italy, [11]RIKEN, Wako, Japan, [12]OATo - INAF Turin, Italy, [13]Russian Space Corporation Energia, Moscow, Russia, [14]National Centre for Nuclear Research, Lodz, Poland, [15]UTIU Rome, Italy, [16]Omega, Ecole Polytechnique, CNRS/IN2P3, Palaiseau, France, [17]INFN - Laboratori Nazionali di Frascati, Italy, [18]Konan University, Japan*

*E-mail:* `bertaina@to.infn.it`, `miyamoto@to.infn.it`

The TurLab facility is a laboratory, equipped with a 5 m diameter and 1 m depth rotating tank, located in the fourth basement level of the Physics Department of the University of Turin. In the past years, we have used the facility to perform experiments related to the observations of Extreme Energy Cosmic Rays (EECRs) from space using the fluorescence technique for JEM-EUSO missions with the main objective to test the response of the trigger logic. In the missions, the diffuse night brightness and artificial and natural light sources can vary significantly in time and space in the Field of View (FoV) of the telescope. Therefore, it is essential to verify the detector performance and test the trigger logic under such an environment.

By means of the tank rotation, a various terrestrial surface with the different optical characteristics such as ocean, land, forest, desert and clouds, as well as artificial and natural light sources such as city lights, lightnings and meteors passing by the detector FoV one after the other is reproduced. The fact that the tank located in a very dark place enables the tests under an optically controlled environment. Using the Mini-EUSO data taken since 2019 onboard the ISS, we will report on the comparison between TurLab and ISS measurements in view of future experiments at TurLab. Moreover, in the forthcoming months we will start testing the trigger logic of the EUSO-SPB2 mission. We report also on the plans and status for this purpose.

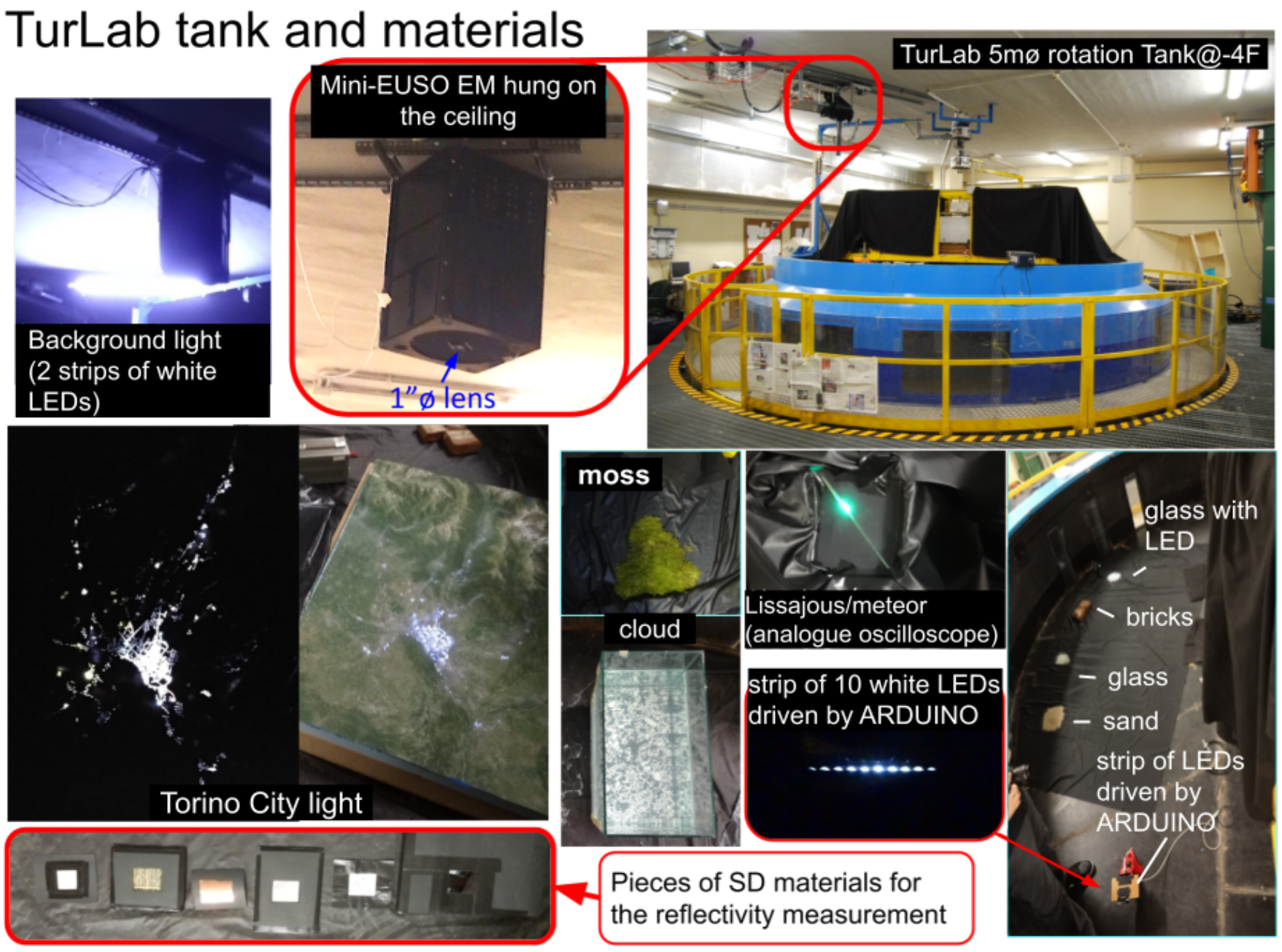

**Figure 1:** Measurement setup and materials to test at the TurLab tank located at -4th floor of Physics department building of University of Turin. Each materials are passing by in the FoV of the telescope as the tank rotates to simulate the Mini-EUSO observation on the ISS orbit.

## 1. Introduction

TurLab [1] is a laboratory for geo-fluid-dynamics studies, where rotation is a key parameter such as Coriolis force and Rossby Number. By using inks or particles, based on the fluid dynamics, key phenomena such as planetary atmospheric and fluid instabilities can be reproduced in the TurLab water tank. The tank has 5 m diameter with a capability of the rotation at a speed of 3 s to 20 min per rotation. Also, as it is located in a very dark environment, the intensity of background light can be adjusted in a controlled condition. The EUSO@TurLab project [2] is a series of measurement campaign, in which we have tested several kinds of prototypes and pathfinders of the fluorescence telescopes equipped with the "EUSO electronics". As described in the following sections, those telescopes are designed to observe the Earth's atmosphere from the stratosphere or space from the orbit of the International Space Station (ISS). By means of the rotating tank with the capabilities mentioned above, we have been testing the EUSO electronics such as its basic performance and the first level trigger (L1) for cosmic-rays, in view of various and even changing background conditions as well as atmospheric phenomena such as meteors and lightning that EUSO telescopes will observe. Mini-EUSO [3] is a scientific mission within the JEM-EUSO program [4]. The telescope has been launched in August 2019 and currently in operation onboard the ISS. The main goal of Mini-EUSO is to measure the UV emissions from the ground and atmosphere, using an orbital platform. These observations will provide interesting data for the scientific study of a variety of UV phenomena such as transient luminous events (TLEs) [7], space debris [8], meteors and hypothetical strange quark matter (SQM) [9] and bioluminescence [5]. Moreover, this will allow us to characterise the UV emission level, which is essential for the optimisation of the design of future EUSO instruments for EECR detection. Mini-EUSO observes the atmosphere from a nadir-facing window inside the Zvezda module of the ISS. It is based on one EUSO detection unit, referred to as the Photo Detector Module (PDM). The PDM consists of 36 multi-anode photomultiplier tubes (MAPMTs), each one having 64 pixels, for a total of 2304 pixels. The MAPMTs are provided by Hamamatsu Photonics, model R11265-M64, and are covered with a 2 mm thickness of BG3 UV filter with anti-reflective coating. The full Mini-EUSO telescope consists of 3 main systems: the optical system, the PDM and the data acquisition system [6]. The optical system of 2 Fresnel lenses is used to focus light onto the PDM in order to achieve a large FoV ($44° \times 44°$) with a relatively light and compact design, well-suited for space application [10]. The PDM detects UV photons and is read out by the data acquisition system with a sampling rate of 2.5 $\mu$s (this is defined as 1 Gate Time Unit, GTU) and a spatial resolution of ~6 km. Data are then processed by a Zynq based FPGA board which implements a multi-level triggering, allowing the measurement of triggered UV transients for 128 frames at time scale of both 2.5 $\mu$s and 320 $\mu$s (D2_GTU). An untriggered acquisition mode with $\simeq$40.96 ms (D3_GTU) frames perms continuous data taking.
The Extreme Universe Space Observatory on a Super Pressure Balloon II (EUSO-SPB2) [11] is an ultra-long-duration balloon mission that aims at the detection of UHECRs via the fluorescence technique and of Very High Energy (VHE) neutrinos via Cherenkov emission. The Fluorescence Telescope [12] consists of 3 PDMs, with 108 MAPMTs, 6912 pixels in total, covering a $37.4 \times 11.4°$ Field of Regard (FoR). The Cherenkov telescope is equipped with a 512-pixel SiPM camera covering a $12.8° \times 6.4°$ FoV. The Fluorescence Telescope looks at the nadir to measure the fluorescence emission from UHECR-induced extensive air shower , while the Cherenkov Telescope is optimised for fast signals (~10 ns) and points near the Earth's limb to detect Cherenkov light from Earth skimming VHE neutrinos or from UHECRs. The mission is planned to fly in 2023 from Wanaka, New Zealand targeting a duration of up to 100 days. Such a flight would provide hundreds of UHECR Cherenkov signals in addition to tens of UHECR fluorescence tracks. It is also a pathfinder mission for instruments of the Probe of Extreme Multi-Messenger Astrophysics (POEMMA) project [13].

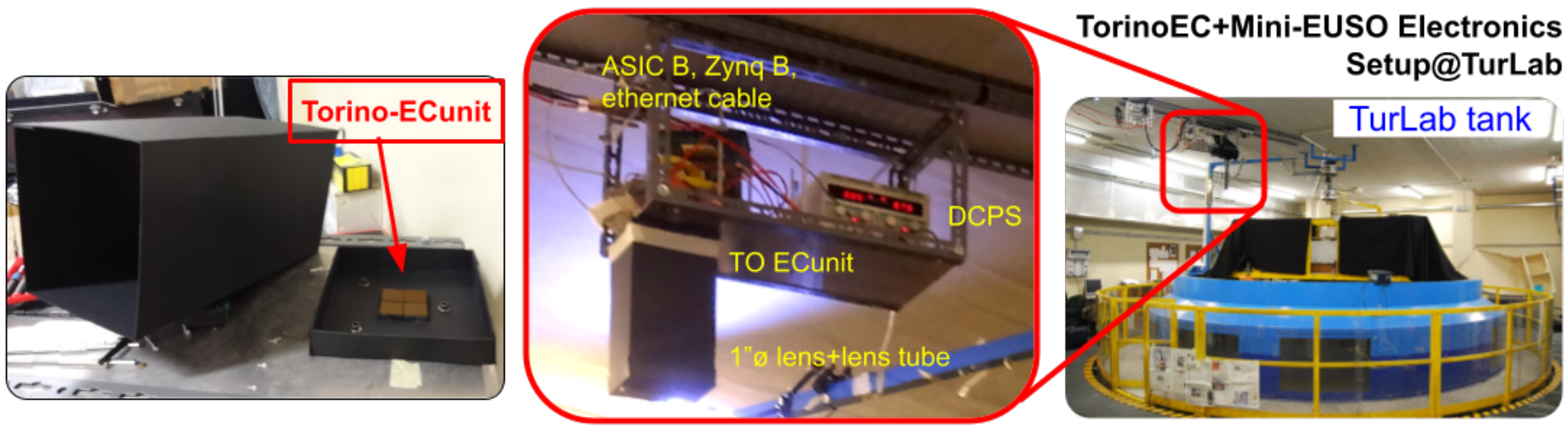

**Figure 2:** Torino EC telescope built in 2019 to test Mini-EUSO trigger. It consists of one ECunit and front-end and data processing system of Mini-EUSO, external high and low voltage power supplies.

## 2. EUSO@TurLab project

The EUSO@TurLab project [2] is a series of measurement campaign, in which we have tested several kinds of prototypes and pathfinders of the fluorescence telescopes equipped with the "EUSO electronics". Those telescopes are designed to observe the Earth's atmosphere from the stratosphere or space from the orbit of the ISS. By means of the rotating tank, we have been testing the EUSO electronics such as its basic performance and the first level trigger (L1) for cosmic-rays, in view of various and even changing background conditions as well as atmospheric phenomena such as meteors and lightning that EUSO telescopes will observe. The Fig. 3 shows the light curve in summed D3 data of a full tank rotation taken by one EC-unit with a Mini-EUSO type electronics. During the year 2018 and 2019, Mini-EUSO Engineering Model (Mini-EUSO EM) [14] and Torino EC telescope assembled with Mini-EUSO electronics are tested at TurLab [15]. Top panel of Fig. 3 shows the summed photon counts of one EC as a function of time in D3_GTU (=40.96 ms) during a full rotation of TurLab measurement, while bottom one shows a summed photon counts but of entire PDM as a function of time (D3_GTU) during the observation for one night on orbit. Also, some atmospheric phenomena such as lightning and meteors are reproduced.

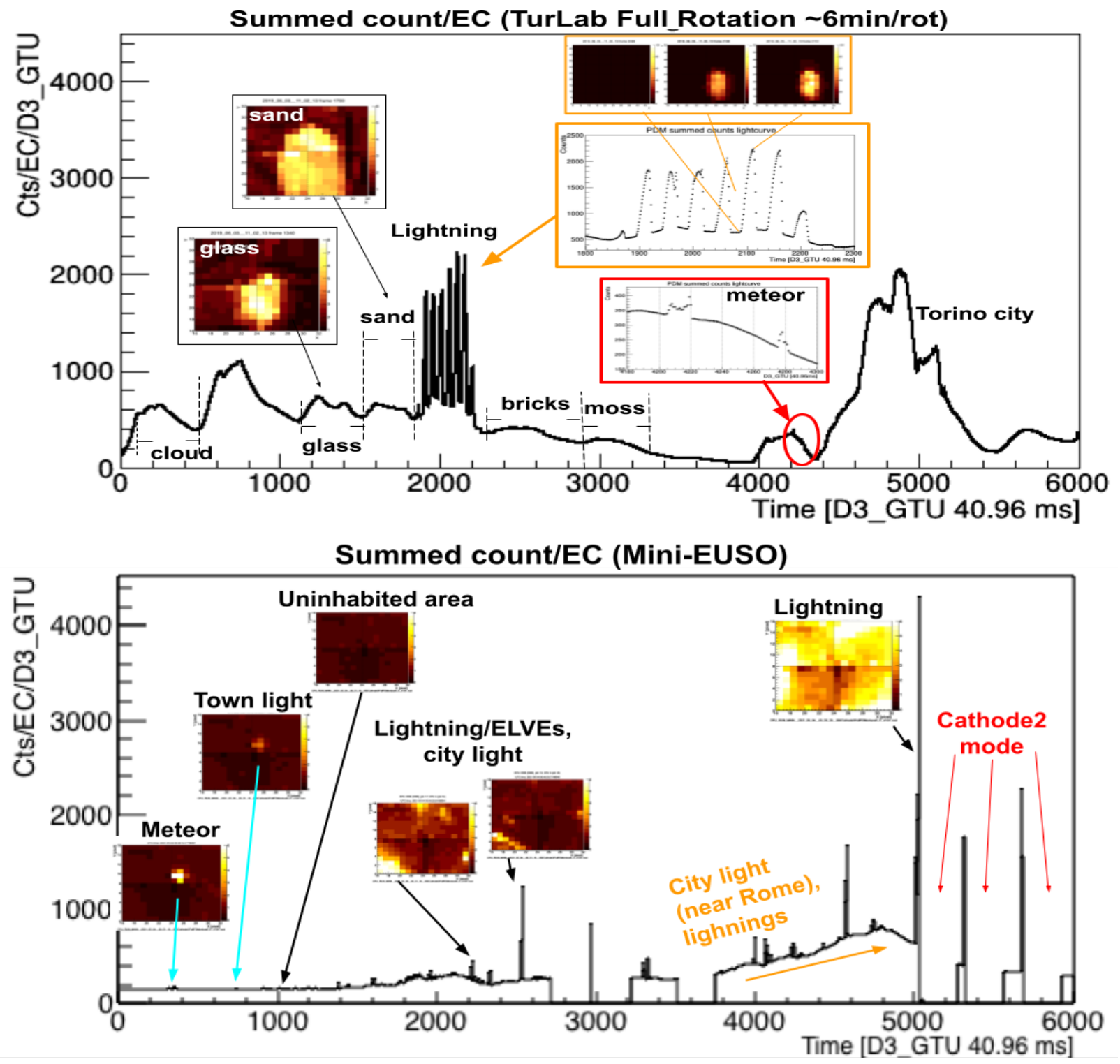

**Figure 3:** Top: summed counts of entire EC (256 pixels) during the tank rotation at a speed of $\sim 6\ min/rot$. TurLab lightning are generated by illuminating ground glass with a white pulse LED. Bottom: an example of Mini-EUSO data in summed counts of entire PDM (2304 pixel) in one night observation in orbit.

### Comparison with Mini-EUSO data

Fig. 4 shows a comparison between an example of meteor event reproduced in TurLab measurement by Lissajous (left) and the meteor events detected by Mini-EUSO (middle and right). In Mini-EUSO data, there are meteors with different brightness and time duration. The meteor we simulated at TurLab has characteristics that have been found among such meteors.

To see the performance of the telescope against a city, we made a scaled size of city of Turin. The scale factor is arbitrary due to the restriction of the size, also it is not our main purpose to reproduce exact image of a particular city but to see the performance against general cities as well as to investigate parameters such as light intensity or density to reproduce the city view in Mini-EUSO. Top part of Fig. 5 shows the representation of Torino city and surroundings. The two images show that villages and cities brighten few pixels or entire MAPMT depending on their extension.

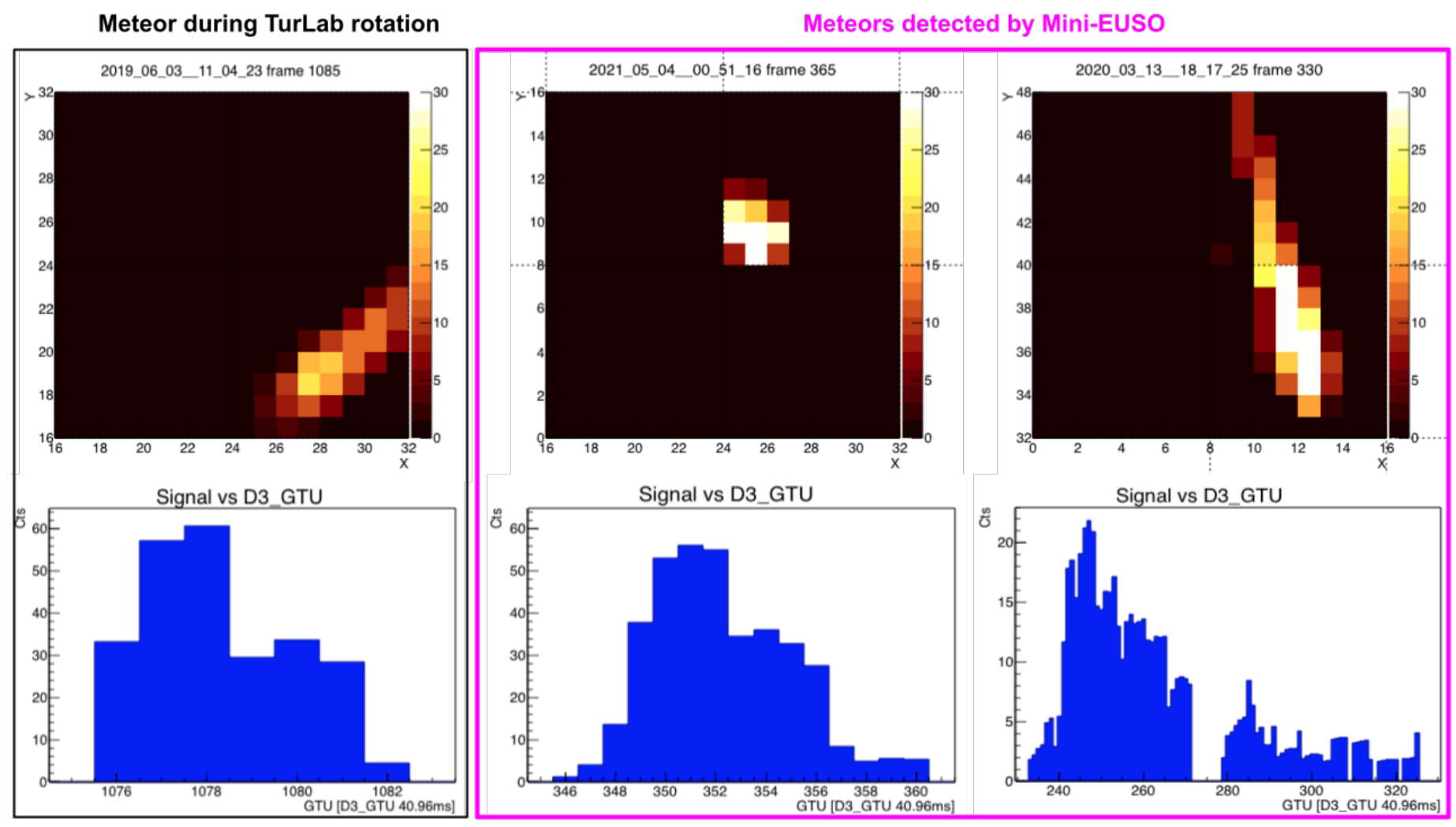

**Figure 4:** Left: The image of a meteor event reproduced in the TurLab measurement during tank rotation which is passing by through a PMT (top) and its time evolution (bottom) as a function of D3_GTU (40.96ms). Middle and Right: Examples of meteor events detected by Mini-EUSO. Integrated images (top) and time evolution plots (bottom). The blank part in the time evolution plot on the right is due to the gap between 2 PMTs where meteor is passing through.

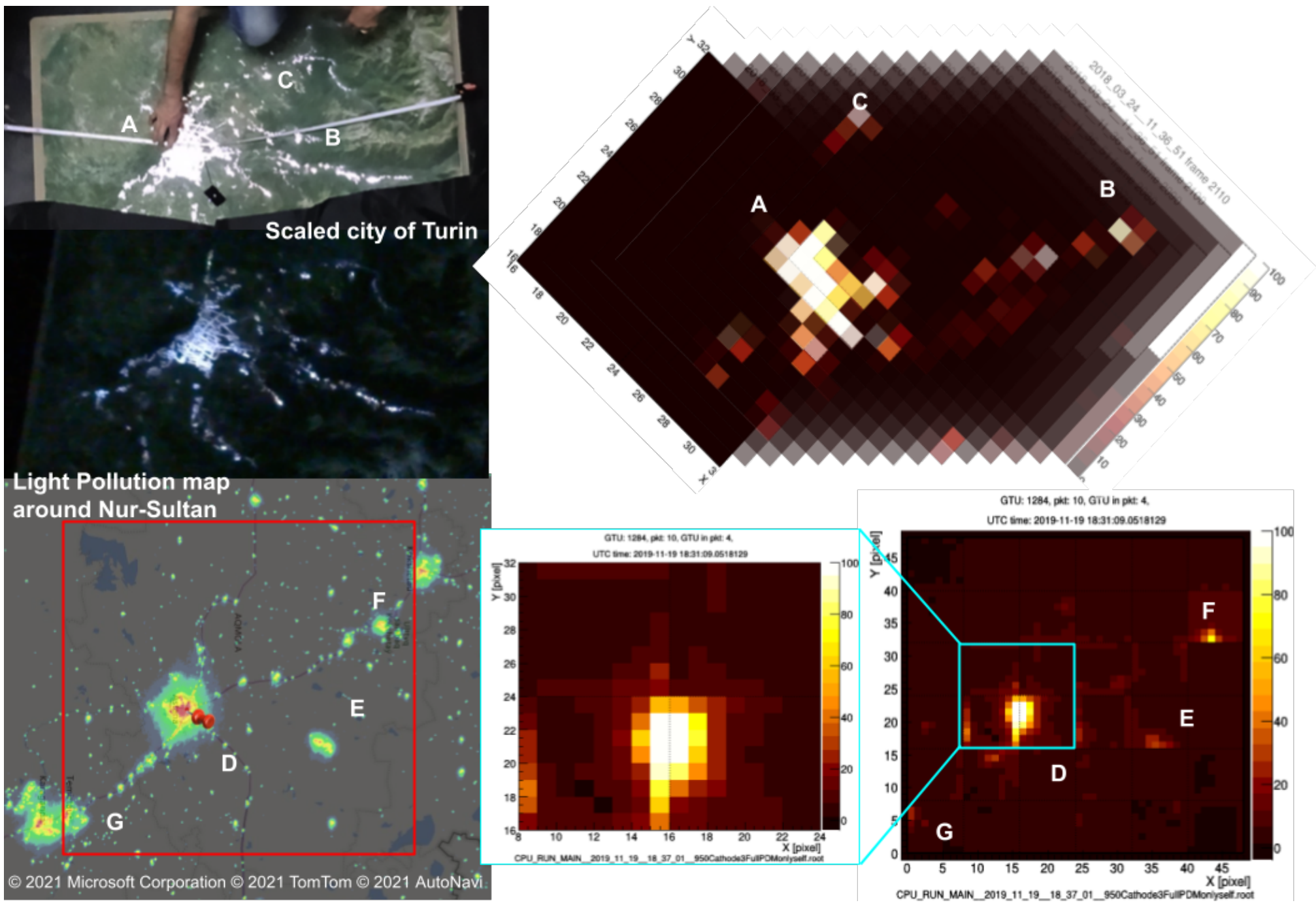

**Figure 5:** Top: Photos of reproduced night light of scaled city of Turin (left), and the images taken by Mini-EUSO Engineering Model in TurLab measurement, superposed every 10 D3_GTUs across the entire city and its suburbs (right). The tank rotation speed for this run corresponds to three times faster than the one corresponding to Mini-EUSO. Bottom: A light pollution map of Nur-Sultan, one of a cities observed by Mini-EUSO (left), and the image taken by Mini-EUSO. Red square indicates Mini-EUSO FoV, each major cities indicated in the pollution map or the photo of Torino city in TurLab with Alphabet correspond to Mini-EUSO PDM image on the right.

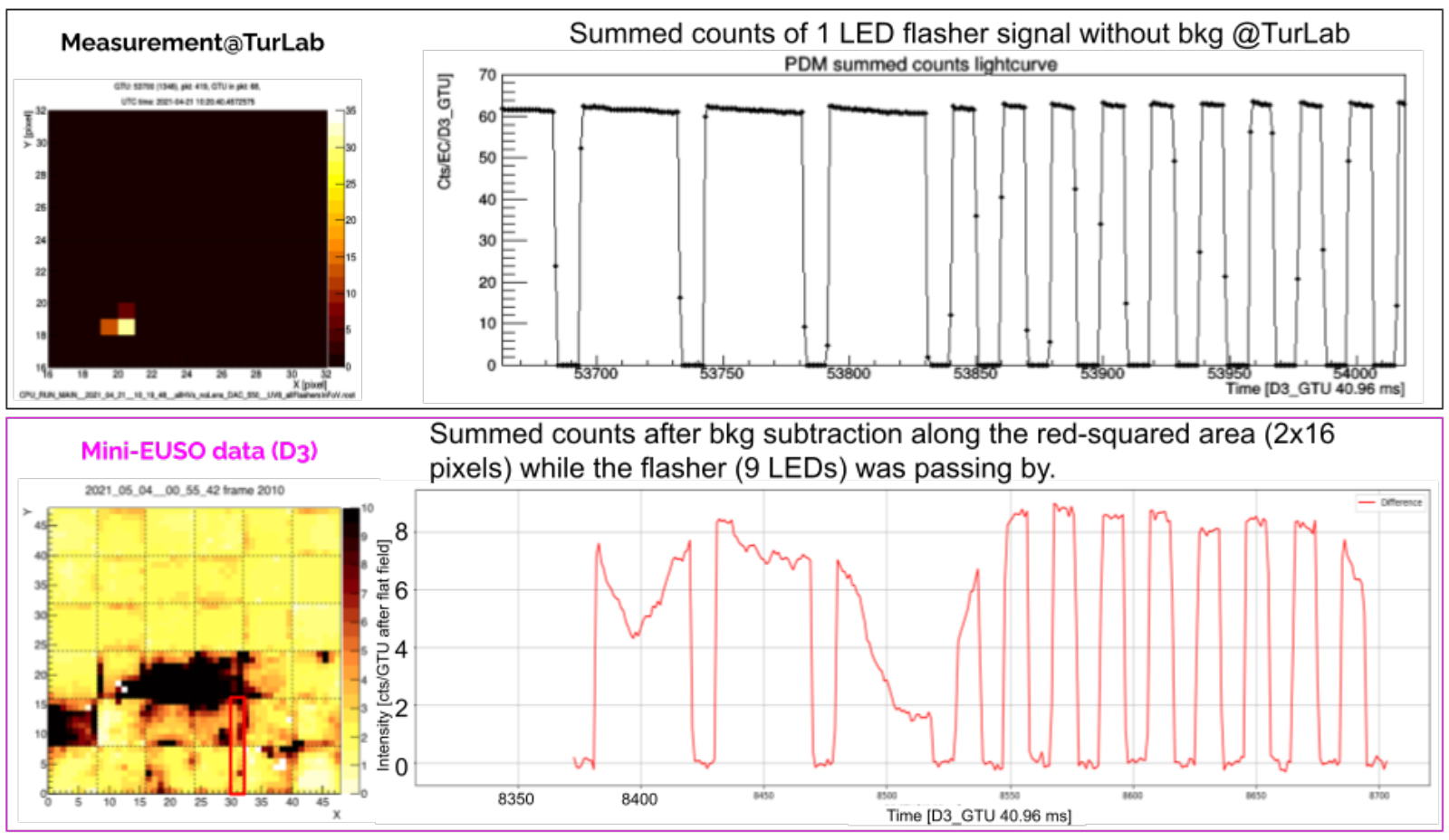

**Figure 6:** Top: Summed counts of 1 LED flasher measured at TurLab. Bottom: Summed counts of 9 LEDs flasher observed by Mini-EUSO.

This situation is compared with images taken by Mini-EUSO, bottom part of Fig.5, on Nur-Sultan (Astana), the capital city of Kazakhstan. From population and altitude point of view, Nur-Sultan is similar to Turin ($\sim 1 million$ for both, located at 347 m and at 239m respectively), however, the area is 3 times bigger. It has also full of golden buildings, enormous shiny objects and even bright laser illuminating the sky for a decoration. Such a city has very similar profile to the representation of Torino city from the point of photon counts and area. In Mini-EUSO FoV the largest urbanised areas extend on EC or PMT scales, while smaller villages appear in groups of 2-4 pixels. This is similar to the transit of Torino map at TurLab. When Torino city crosses the FoV of the camera half EC is illuminated while when other small villages cross the FoV only few pixels are brightened.

## 3. Measurement@TurLab for a Flasher campaign May 2021

Several kinds of ground based flashers have been developed by some groups of Mini-EUSO collaboration such as in Japan, Italy and Russia. One of such flashers has been developed in Turin and tested in advance with the Torino EC telescope at the TurLab facility, where the distance of 40 m between the telescope and the light source is available in the dark environment at -4th floor. The flasher consists of an array of 9 100W COB-UV LEDs, DC power supply and Arduino circuit. Taking into account that Mini-EUSO pixel FoV is ~6 km passing by at a velocity of ISS which is ~7.5 km/s, it will take 800 ms to pass completely one pixel. Thus, to be sure one pixel has constant full pulse, and also to see the transit of a pulse within a pixel, LEDs were pulsed 6 times in 12 s with a pulse of 1600 ms on and 400 ms off each, followed by 12 pulses in 9.6 s with 400 ms on and 400 ms off each. To reduce the light as well as to obtain only parallel light, we collimated the light at 30 cm distance from the detector focal surface with a pin-hole of 0.1 mm diameter. As a result, the total number of photons we obtained is ~60 cts/LED, which corresponds to 87.7 cts/pix/GTU with pile-up correction for each LEDs as shown in the top panel of Fig. 6. As described above, the MAPMTs and UV filter, as well as the electronics used for the measurements on the ground and orbit are the same type. Taking into account the ratio of parameters between the two telescopes, such as distance, optics, signal absorption by atmosphere, transmittance of UV window on the ISS, incident position and angle dependence of the Mini-EUSO optics, number of the LEDs (1 for the ground, 9 for the orbit) the detected number is well matching to the expected number to be detected (see bottom panels of Fig. 6) by Mini-EUSO. We will calibrate either or both of the telescope on the ground or flasher LEDs in the future to achieve an end-to-end calibration of Mini-EUSO.

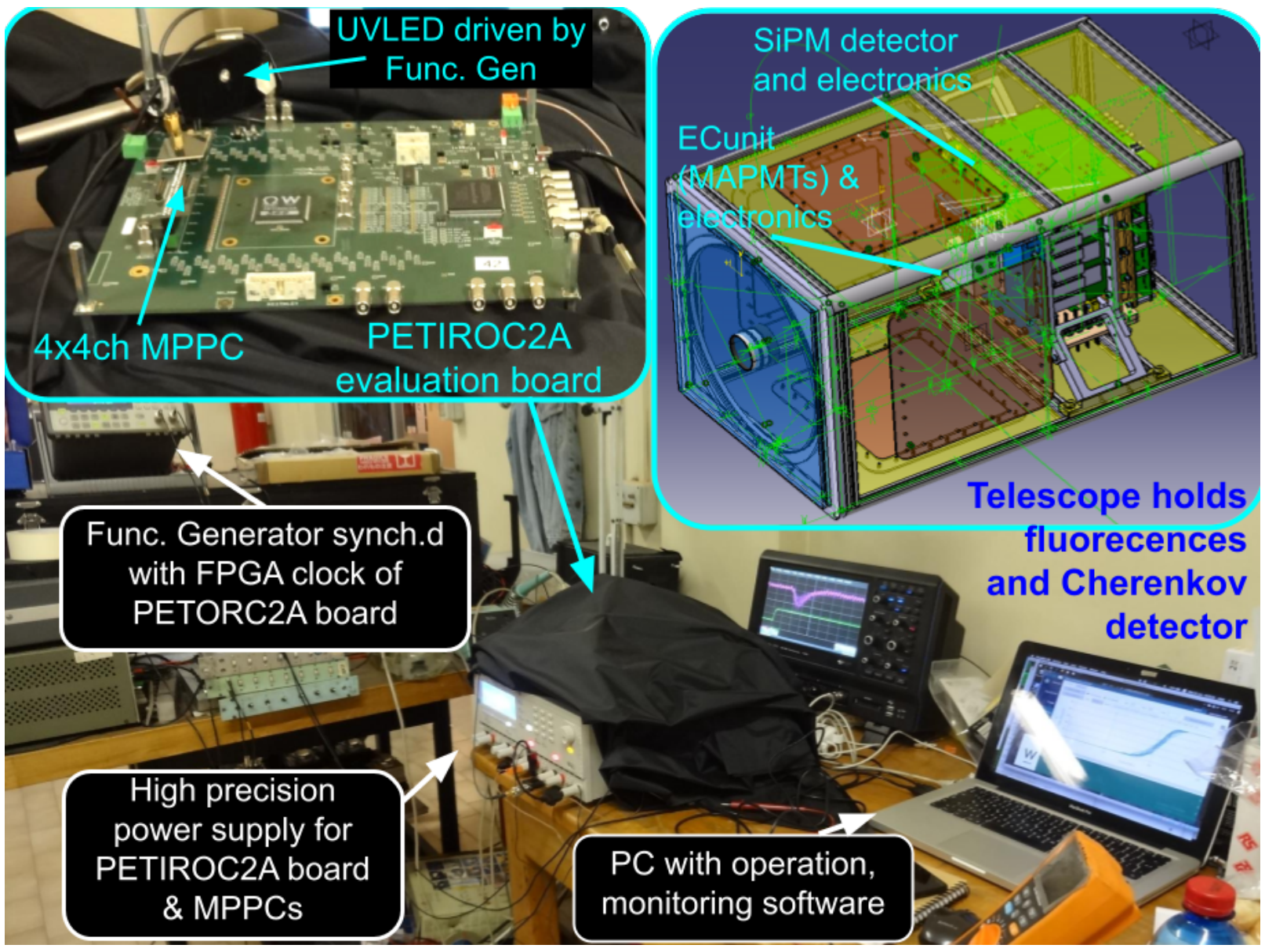

**Figure 7:** Preparation for the TurLab measurement for EUSO-SPB2 type detector. Using a 16-ch SiPM device and its dedicated test board (left-top), functional tests are ongoing as shown in the bottom as an example of measurement setup for the test. In parallel, a telescope which can hold both Fluorescence telescope-like system (1 ECunit consisting of 4 MAPMTs and electronics) and Cherenkov telescope-like system (SiPM and its electronics) is designed and going to be built for the TurLab measurement as well as open-sky observation to understand its fundamental performance as well as to test the trigger algorithm.

## 4. Outlook

Similar kind of measurement is planned to be done for EUSO-SPB2 detector. Main EUSO-SPB2 detectors consists of fluorescence telescope and Cherenkov telescope. Each of 3 PDMs of the fluorescence telescope consists of 36 64ch-MAPMTs like Mini-EUSO one but different trigger system. The Cherenkov detector consists of SiPMs and dedicated front-end electronics. As shown in Fig. 7, we're currently building a telescope holding a prototype of fluorescence detector and SiPM array and their dedicated electronics to test its performance and trigger algorithm at TurLab. The prototype of fluorescence telescope consists of EUSO-SPB2 electronics and HVPS and one EC-unit instead of 9. For the SiPM detector, we use one or two Hamamatsu s13361-6050AS-04 MPPCs and its evaluation board to understand its performance against various kinds of background and to test the trigger algorithm.

## 5. Conclusions

The EUSO@TurLab project is an ongoing activity with the aim of reproducing the luminous conditions in a laboratory environment that a project of the JEM-EUSO program will see while it's flying in space or on stratospheric balloon platforms. Mini-EUSO type telescope as well as Mini-EUSO Engineering Model are tested in the TurLab facility and the data are compared with the Mini-EUSO data. Further analysis is currently ongoing. The preparation for testing EUSO-SPB2 type of detector is also currently ongoing and we will test it in forthcoming months toward the lunch of EUSO-SPB2 detector.

## 6. Acknowledgements

This work was supported by State Space Corporation ROSCOSMOS, by the Italian Space Agency through the ASI INFN agreement n. 2020-26-HH.0 and contract n. 2016-1-U.0, by the French Space agency CNES, National Science Centre in Poland grant 2017/27/B/ST9/02162. This research has been supported by the Interdisciplinary Scientific and Educational School of Moscow University "Fundamental and Applied Space Research". The article has been prepared based on research materials carried out in the space experiment "UV atmosphere". The authors express their deep and collegial thanks to the entire JEM-EUSO program and all its individual members.

## Full Authors List: JEM-EUSO Collaboration

G. Abdellaoui[ah], S. Abe[fq], J.H. Adams Jr.[pd], D. Allard[cb], G. Alonso[md], L. Anchordoqui[pe], A. Anzalone[eh,ed], E. Arnone[ek,el], K. Asano[fe], R. Attallah[ac], H. Attoui[aa], M. Ave Pernas[mc], M. Bagheri[ph], J. Baláz[la], M. Bakiri[aa], D. Barghini[el,ek], S. Bartocci[ei,ej], M. Battisti[ek,el], J. Bayer[dd], B. Beldjilali[ah], T. Belenguer[mb], N. Belkhalfa[aa], R. Bellotti[ea,eb], A.A. Belov[kb], K. Benmessai[aa], M. Bertaina[ek,el], P.F. Bertone[pf], P.L. Biermann[db], F. Bisconti[el,ek], C. Blaksley[ft], N. Blanc[oa], S. Blin-Bondil[ca,cb], P. Bobik[la], M. Bogomilov[ba], K. Bolmgren[na], E. Bozzo[ob], S. Briz[pb], A. Bruno[eh,ed], K.S. Caballero[hd], F. Cafagna[ea], G. Cambié[ei,ej], D. Campana[ef], J-N. Capdevielle[cb], F. Capel[de], A. Caramete[ja], L. Caramete[ja], P. Carlson[na], R. Caruso[ec,ed], M. Casolino[ft,ei], C. Cassardo[ek,el], A. Castellina[ek,em], O. Catalano[eh,ed], A. Cellino[ek,em], K. Černý[bb], M. Chikawa[fc], G. Chiritoi[ja], M.J. Christl[pf], R. Colalillo[ef,eg], L. Conti[en,ei], G. Cotto[ek,el], H.J. Crawford[pa], R. Cremonini[el], A. Creusot[cb], A. de Castro Gónzalez[pb], C. de la Taille[ca], L. del Peral[mc], A. Diaz Damian[cc], R. Diesing[pb], P. Dinaucourt[ca], A. Djakonow[ia], T. Djemil[ac], A. Ebersoldt[db], T. Ebisuzaki[ft], J. Eser[pb], F. Fenu[ek,el], S. Fernández-González[ma], S. Ferrarese[ek,el], G. Filippatos[pc], W.I. Finch[pc] C. Fornaro[en,ei], M. Fouka[ab], A. Franceschi[ee], S. Franchini[md], C. Fuglesang[na], T. Fujii[fg], M. Fukushima[fe], P. Galeotti[ek,el], E. García-Ortega[ma], D. Gardiol[ek,em], G.K. Garipov[kb], E. Gascón[ma], E. Gazda[ph], J. Genci[lb], A. Golzio[ek,el], C. González Alvarado[mb], P. Gorodetzky[ft], A. Green[pc], F. Guarino[ef,eg], C. Guépin[pl], A. Guzmán[dd], Y. Hachisu[ft], A. Haungs[db], J. Hernández Carretero[mc], L. Hulett[pc], D. Ikeda[fe], N. Inoue[fn], S. Inoue[ft], F. Isgrò[ef,eg], Y. Itow[fk], T. Jammer[dc], S. Jeong[gb], E. Joven[me], E.G. Judd[pa], J. Jochum[dc], F. Kajino[ff], T. Kajino[fi], S. Kalli[af], I. Kaneko[ft], Y. Karadzhov[ba], M. Kasztelan[ia], K. Katahira[ft], K. Kawai[ft], Y. Kawasaki[ft], A. Kedadra[aa], H. Khales[aa], B.A. Khrenov[kb], Jeong-Sook Kim[ga], Soon-Wook Kim[ga], M. Kleifges[db], P.A. Klimov[kb], D. Kolev[ba], I. Kreykenbohm[da], J.F. Krizmanic[pf,pk], K. Królik[ia], V. Kungel[pc], Y. Kurihara[fs], A. Kusenko[fr,pe], E. Kuznetsov[pd], H. Lahmar[aa], F. Lakhdari[ag], J. Licandro[me], L. López Campano[ma], F. López Martínez[pb], S. Mackovjak[la], M. Mahdi[aa], D. Mandát[bc], M. Manfrin[ek,el], L. Marcelli[ei], J.L. Marcos[ma], W. Marszał[ia], Y. Martín[me], O. Martinez[hc], K. Mase[fa], R. Matev[ba], J.N. Matthews[pg], N. Mebarki[ad], G. Medina-Tanco[ha], A. Menshikov[db], A. Merino[ma], M. Mese[ef,eg], J. Meseguer[md], S.S. Meyer[pb], J. Mimouni[ad], H. Miyamoto[ek,el], Y. Mizumoto[fi], A. Monaco[ea,eb], J.A. Morales de los Ríos[mc], M. Mastafa[pd], S. Nagataki[ft], S. Naitamor[ab], T. Napolitano[ee], J. M. Nachtman[pi] A. Neronov[ob,cb], K. Nomoto[fr], T. Nonaka[fe], T. Ogawa[ft], S. Ogio[fl], H. Ohmori[ft], A.V. Olinto[pb], Y. Onel[pi] G. Osteria[ef], A.N. Otte[ph], A. Pagliaro[eh,ed], W. Painter[db], M.I. Panasyuk[kb], B. Panico[ef], E. Parizot[cb], I.H. Park[gb], B. Pastircak[la], T. Paul[pe], M. Pech[bb], I. Pérez-Grande[md], F. Perfetto[ef], T. Peter[oc], P. Picozza[ei,ej,ft], S. Pindado[md], L.W. Piotrowski[ib], S. Piraino[dd], Z. Plebaniak[ek,el,ia], A. Pollini[oa], E.M. Popescu[ja], R. Prevete[ef,eg], G. Prévôt[cb], H. Prieto[mc], M. Przybylak[ia], G. Puehlhofer[dd], M. Putis[la], P. Reardon[pd], M.H.. Reno[pi], M. Reyes[me], M. Ricci[ee], M.D. Rodríguez Frías[mc], O.F. Romero Matamala[ph], F. Ronga[ee], M.D. Sabau[mb], G. Saccá[ec,ed], G. Sáez Cano[mc], H. Sagawa[fe], Z. Sahnoune[ab], A. Saito[fg], N. Sakaki[ft], H. Salazar[hc], J.C. Sanchez Balanzar[ha], J.L. Sánchez[ma], A. Santangelo[dd], A. Sanz-Andrés[md], M. Sanz Palomino[mb], O.A. Saprykin[kc], F. Sarazin[pc], M. Sato[fo], A. Scagliola[ea,eb], T. Schanz[dd], H. Schieler[db], P. Schovánek[bc], V. Scotti[ef,eg], M. Serra[me], S.A. Sharakin[kb], H.M. Shimizu[fj], K. Shinozaki[ia], J.F. Soriano[pe], A. Sotgiu[ei,ej], I. Stan[ja], I. Strharský[la], N. Sugiyama[fj], D. Supanitsky[ha], M. Suzuki[fm], J. Szabelski[ia], N. Tajima[ft], T. Tajima[ft], Y. Takahashi[fo], M. Takeda[fe], Y. Takizawa[ft], M.C. Talai[ac], Y. Tameda[fp], C. Tenzer[dd], S.B. Thomas[pg], O. Tibolla[he], L.G. Tkachev[ka], T. Tomida[fh], N. Tone[ft], S. Toscano[ob], M. Traïche[aa], Y. Tsunesada[fl], K. Tsuno[ft], S. Turriziani[ft], Y. Uchihori[fb],

O. Vaduvescu[me], J.F. Valdés-Galicia[ha], P. Vallania[ek,em], L. Valore[ef,eg], G. Vankova-Kirilova[ba], T. M. Venters[pj], C. Vigorito[ek,el], L. Villaseñor[hb], B. Vlcek[mc], P. von Ballmoos[cc], M. Vrabel[lb], S. Wada[ft], J. Watanabe[fi], J. Watts Jr.[pd], R. Weigand Muñoz[ma], A. Weindl[db], L. Wiencke[pc], M. Wille[da], J. Wilms[da], D. Winn[pm] T. Yamamoto[ff], J. Yang[gb], H. Yano[fm], I.V. Yashin[kb], D. Yonetoku[fd], S. Yoshida[fa], R. Young[pf], I.S Zgura[ja], M.Yu. Zotov[kb], A. Zuccaro Marchi[ft]

[aa] Centre for Development of Advanced Technologies (CDTA), Algiers, Algeria
[ab] Dep. Astronomy, Centre Res. Astronomy, Astrophysics and Geophysics (CRAAG), Algiers, Algeria
[ac] LPR at Dept. of Physics, Faculty of Sciences, University Badji Mokhtar, Annaba, Algeria
[ad] Lab. of Math. and Sub-Atomic Phys. (LPMPS), Univ. Constantine I, Constantine, Algeria
[af] Department of Physics, Faculty of Sciences, University of M'sila, M'sila, Algeria
[ag] Research Unit on Optics and Photonics, UROP-CDTA, Sétif, Algeria
[ah] Telecom Lab., Faculty of Technology, University Abou Bekr Belkaid, Tlemcen, Algeria
[ba] St. Kliment Ohridski University of Sofia, Bulgaria
[bb] Joint Laboratory of Optics, Faculty of Science, Palacký University, Olomouc, Czech Republic
[bc] Institute of Physics of the Czech Academy of Sciences, Prague, Czech Republic
[ca] Omega, Ecole Polytechnique, CNRS/IN2P3, Palaiseau, France
[cb] Université de Paris, CNRS, AstroParticule et Cosmologie, F-75013 Paris, France
[cc] IRAP, Université de Toulouse, CNRS, Toulouse, France
[da] ECAP, University of Erlangen-Nuremberg, Germany
[db] Karlsruhe Institute of Technology (KIT), Germany
[dc] Experimental Physics Institute, Kepler Center, University of Tübingen, Germany
[dd] Institute for Astronomy and Astrophysics, Kepler Center, University of Tübingen, Germany
[de] Technical University of Munich, Munich, Germany
[ea] Istituto Nazionale di Fisica Nucleare - Sezione di Bari, Italy
[eb] Universita' degli Studi di Bari Aldo Moro and INFN - Sezione di Bari, Italy
[ec] Dipartimento di Fisica e Astronomia "Ettore Majorana", Universita' di Catania, Italy
[ed] Istituto Nazionale di Fisica Nucleare - Sezione di Catania, Italy
[ee] Istituto Nazionale di Fisica Nucleare - Laboratori Nazionali di Frascati, Italy
[ef] Istituto Nazionale di Fisica Nucleare - Sezione di Napoli, Italy
[eg] Universita' di Napoli Federico II - Dipartimento di Fisica "Ettore Pancini", Italy
[eh] INAF - Istituto di Astrofisica Spaziale e Fisica Cosmica di Palermo, Italy
[ei] Istituto Nazionale di Fisica Nucleare - Sezione di Roma Tor Vergata, Italy
[ej] Universita' di Roma Tor Vergata - Dipartimento di Fisica, Roma, Italy
[ek] Istituto Nazionale di Fisica Nucleare - Sezione di Torino, Italy
[el] Dipartimento di Fisica, Universita' di Torino, Italy
[em] Osservatorio Astrofisico di Torino, Istituto Nazionale di Astrofisica, Italy
[en] Uninettuno University, Rome, Italy
[fa] Chiba University, Chiba, Japan
[fb] National Institutes for Quantum and Radiological Science and Technology (QST), Chiba, Japan
[fc] Kindai University, Higashi-Osaka, Japan
[fd] Kanazawa University, Kanazawa, Japan
[fe] Institute for Cosmic Ray Research, University of Tokyo, Kashiwa, Japan

[ff] Konan University, Kobe, Japan
[fg] Kyoto University, Kyoto, Japan
[fh] Shinshu University, Nagano, Japan
[fi] National Astronomical Observatory, Mitaka, Japan
[fj] Nagoya University, Nagoya, Japan
[fk] Institute for Space-Earth Environmental Research, Nagoya University, Nagoya, Japan
[fl] Graduate School of Science, Osaka City University, Japan
[fm] Institute of Space and Astronautical Science/JAXA, Sagamihara, Japan
[fn] Saitama University, Saitama, Japan
[fo] Hokkaido University, Sapporo, Japan
[fp] Osaka Electro-Communication University, Neyagawa, Japan
[fq] Nihon University Chiyoda, Tokyo, Japan
[fr] University of Tokyo, Tokyo, Japan
[fs] High Energy Accelerator Research Organization (KEK), Tsukuba, Japan
[ft] RIKEN, Wako, Japan
[ga] Korea Astronomy and Space Science Institute (KASI), Daejeon, Republic of Korea
[gb] Sungkyunkwan University, Seoul, Republic of Korea
[ha] Universidad Nacional Autónoma de México (UNAM), Mexico
[hb] Universidad Michoacana de San Nicolas de Hidalgo (UMSNH), Morelia, Mexico
[hc] Benemérita Universidad Autónoma de Puebla (BUAP), Mexico
[hd] Universidad Autónoma de Chiapas (UNACH), Chiapas, Mexico
[he] Centro Mesoamericano de Física Teórica (MCTP), Mexico
[ia] National Centre for Nuclear Research, Lodz, Poland
[ib] Faculty of Physics, University of Warsaw, Poland
[ja] Institute of Space Science ISS, Magurele, Romania
[ka] Joint Institute for Nuclear Research, Dubna, Russia
[kb] Skobeltsyn Institute of Nuclear Physics, Lomonosov Moscow State University, Russia
[kc] Space Regatta Consortium, Korolev, Russia
[la] Institute of Experimental Physics, Kosice, Slovakia
[lb] Technical University Kosice (TUKE), Kosice, Slovakia
[ma] Universidad de León (ULE), León, Spain
[mb] Instituto Nacional de Técnica Aeroespacial (INTA), Madrid, Spain
[mc] Universidad de Alcalá (UAH), Madrid, Spain
[md] Universidad Politécnia de madrid (UPM), Madrid, Spain
[me] Instituto de Astrofísica de Canarias (IAC), Tenerife, Spain
[na] KTH Royal Institute of Technology, Stockholm, Sweden
[oa] Swiss Center for Electronics and Microtechnology (CSEM), Neuchâtel, Switzerland
[ob] ISDC Data Centre for Astrophysics, Versoix, Switzerland
[oc] Institute for Atmospheric and Climate Science, ETH Zürich, Switzerland
[pa] Space Science Laboratory, University of California, Berkeley, CA, USA
[pb] University of Chicago, IL, USA
[pc] Colorado School of Mines, Golden, CO, USA
[pd] University of Alabama in Huntsville, Huntsville, AL; USA

[pe] Lehman College, City University of New York (CUNY), NY, USA
[pf] NASA Marshall Space Flight Center, Huntsville, AL, USA
[pg] University of Utah, Salt Lake City, UT, USA
[ph] Georgia Institute of Technology, USA
[pi] University of Iowa, Iowa City, IA, USA
[pj] NASA Goddard Space Flight Center, Greenbelt, MD, USA
[pk] Center for Space Science & Technology, University of Maryland, Baltimore County, Baltimore, MD, USA
[pl] Department of Astronomy, University of Maryland, College Park, MD, USA
[pm] Fairfield University, Fairfield, CT, USA